\documentstyle[12pt,worldsci]{article}
\def\g{\gamma}
\def\d{\delta}

\def\L{\Lambda}

\def\ha{\frac{1}{2}}
\def\bar#1{\overline{ #1 }}
\def\psibar{\overline{\psi}}

\def\beq{\begin{equation}}
\def\eeq{\end{equation}}
\def\bea{\begin{eqnarray}}
\def\eea{\end{eqnarray}}
\def\bit{\begin{itemize}}
\def\eit{\end{itemize}}
\def\ben{\begin{enumerate}}
\def\een{\end{enumerate}}

\def\bdel{{\bf \nabla}}
\def\bA{{\bf A}}

\def\bD{{\bf D}}

\def\bx{{\bf x}}
\def\by{{\bf y}}

\def\bPi{{\bf \Pi}}

\def\FP{{\cal J}}
\def\pri{\prime}
\def\Heff{H_{\rm eff}}
\def\Hcan{H_{\rm can}}

\begin{document}
\title{RENORMALIZED EFFECTIVE HAMILTONIAN APPROACH TO QCD}
\author{D. G. Robertson,$^1$ E. S. Swanson,$^1$ A. P.
Szczepaniak,$^2$
C.-R. Ji$^1$ and S. R. Cotanch$^1$\\
{\em $^1$Department of Physics, North Carolina State University,
Raleigh, NC 27695, USA}\\
{\em $^2$Department of Physics and Nuclear Theory Center, Indiana
University, Bloomington, IN 47405, USA}}
\maketitle
\setlength{\baselineskip}{2.6ex}

\vspace{0.7cm}
\begin{abstract}
Continuing our previous QCD Hamiltonian studies in the gluonic and
quark sectors, we describe a new renormalization procedure which
generates an effective Hamiltonian.  The formulation, which is in the
Coulomb gauge, provides an improved framework for investigating hadron
structure.
\end{abstract}
\vspace{0.7cm}

\section{Introduction}

For the past few years the NCSU group has focused upon the development
of a comprehensive relativistic, many-body description of hadron
structure {\cite{bigpaper,ncsu}}.  This approach has many virtues,
particularly the explicit incorporation of gluonic degrees of freedom
and a rigorous connection to QCD at short distances.  Calculations
{\cite{ncsu}} of the glueball spectrum have been quite encouraging,
reproducing the available lattice data.

This talk details recent results from a new renormalization procedure
for the Coulomb gauge QCD Hamiltonian.  There are several reasons for
selecting the Coulomb gauge.  First, it is completely physical; there
are no spurious gauge degrees of freedom or ghosts.  Second, both
$q\bar q$ and gluon-gluon interactions appear explicitly in the
canonical Hamiltonian which facilitates contact with quark model
phenomenology. Finally, there appears to be a natural scenario for
confinement in the Coulomb gauge, as originally suggested by Gribov
{\cite {gribov}} and recently formalized by Zwanziger
{\cite{zwanziger}}.

This Hamiltonian must be supplemented with a regulator, which is
applicable nonperturbatively and is consistent with the many-body
techniques we eventually wish to employ.  Next the renormalized
Hamiltonian is constructed, which includes the correct cut-off
dependent counterterms such that physical quantities (e.g.,
eigenvalues) are cut-off independent.  It is inevitable that the
regulators at our disposal will violate subsets of Lorentz covariance
and gauge invariance, so that the structure of the counterterm
Hamiltonian may be expected to be quite complicated.  Due to
asymptotic freedom, however, it should be reasonable to construct
these operators using perturbation theory, provided the energy cut-off
is not too low.

The result is a Hamiltonian formulation of QCD which is fully
covariant and gauge invariant through some order in perturbation
theory.  To effectively apply this Hamiltonian, further approximations
and phenomenological considerations are necessary.  An approach we
have initiated involves making a variational {\em Ansatz} for the
physical vacuum state, based on the BCS ``pairing'' form.
Quasi-particle degrees of freedom emerge naturally with
constituent-scale masses and may be identified with constituent quarks
and gluons.  One then can construct approximate bound states of these
constituents utilizing standard many-body techniques such as
Tamm-Dancoff and RPA.

\section{Regularization and Renormalization}

Our starting point is the canonical QCD Hamiltonian in the Coulomb
gauge $\bdel\cdot\bA=0$.  The dynamical degrees of freedom are the
transverse gauge fields $\bA$, their conjugate momenta $\bPi$, and the
quark field $\psi$.  The canonical Hamiltonian may be written as
\begin{equation}
\Hcan = H_0 + H_{gg} + H_{qg} + H_C \; ,
\label{breakup}
\end{equation}
where $H_0$ is the free Hamiltonian for quarks and transverse gluons,
$H_{gg}$ contains the self-interactions between transverse gluons, and
$H_{qg}$ contains quark-gluon interactions.  $H_C$ is the Coulomb
term,
\begin{equation}
H_C = \ha g^2 \int d^3x d^3y\thinspace
\FP^{-1}\rho^a(\bx) K^{ab}(\bx,\by) \FP \rho^b(\by)\; ,
\label{coulomb}
\end{equation}
where $\FP= {\rm det}\left[\bdel\cdot\bD\right]$ is the Fadeev-Popov
determinant, and the kernel $K$ is given by
\begin{equation}
K^{ab}(\bx,\by) = \langle\bx,a|(\bdel\cdot\bD)^{-1}(-\bdel^2)
(\bdel\cdot\bD)^{-1}|\by,b\rangle\; .
\end{equation}
The color charge density $\rho^a$ includes both matter and gluonic
contributions, $\rho^a(\bx) = \psi^\dagger(\bx) T^a\psi(\bx) +f^{abc}
\bA^b(\bx) \cdot \bPi^c(\bx)$, so that $H_C$ represents a complicated
set of instantaneous quark-quark, quark-gluon, and gluon-gluon
interactions.

Perturbative schemes for renormalizing Hamiltonians typically suffer
from technical problems related to the occurrence of vanishing energy
denominators.  One way to avoiding this difficulty is the cut-off
method of G{\l}azek and Wilson {\cite{gw93}}.  In this approach one
considers the basis formed by eigenstates of the free Hamiltonian
$H_0$, with eigenvalues $E_n$.  The theory is then regulated by
suppressing matrix elements of $H$ between states for which the
absolute value of the free energy difference $E_{nm}\equiv E_n-E_m$ is
large.  Specifically, we define matrix elements of the regulated
Hamiltonian $H(\L)$ by
\begin{equation}
\langle n|H(\L)|m \rangle \equiv E_n \d_{nm}
+f_{mn}(\L) \langle n|H_I |m\rangle\; ,
\end{equation}
where $H_I=H-H_0$ is the interaction.  Here $f_{nm}(\L)$ is some
convenient function which is unity for $|E_{nm}|\ll \L$ and vanishes
for $|E_{nm}|\gg \L$.  In Ref.  {\cite{bigpaper}} we chose
\begin{equation}
f_{nm}(\L) = e^{-E_{nm}^2/\L^2}\; .
\label{cut-off}
\end{equation}
Note that, since the cut-off is defined in terms of {\em free} energy
differences, it is not fully Lorentz covariant (though it is of course
rotationally invariant).  It also violates gauge invariance, though
this issue is difficult to address directly in the present context
since we work in a fixed gauge.  The counterterm Hamiltonian will
therefore also contain Lorentz- and gauge-noninvariant terms, which
are necessary to correct for the violations induced by the regulator.
These complications are in practice unavoidable if one wishes to work
with Hamiltonians in a relativistic field theory.

At this point we have a fully regulated formulation of QCD in the
Coulomb gauge.  Our next task is to remove the dependence on the
cut-off parameter $\L$ by adding counterterms to the Hamiltonian.  We
thus obtain the renormalized effective Hamiltonian, $\Heff(\L)$, which
may then be analyzed nonperturbatively using many-body techniques.

An elegant way of determining the counterterms involves designing a
similarity transformation which changes the cut-off in all matrix
elements of $H$.  Because the transformed Hamiltonian is equivalent to
the original one, physical predictions (e.g., eigenvalues) are
unchanged.  One then searches for Hamiltonians which are form
invariant, or ``coherent,'' under this transformation.  Such
self-similar Hamiltonians will automatically yield cut-off independent
predictions and thus represent renormalized Hamiltonians.
Specifically, we write
\begin{equation}
H(\Lambda^\pri) = U(\L^\pri;\L) H(\Lambda) U^\dagger(\L^\pri;\L)\; ,
\label{simtrans}
\end{equation}
where $U$ is a unitary matrix constructed so that the matrix elements
$\langle m|H(\Lambda^\pri)|n \rangle$ are all proportional to
$\exp(-E_{nm}^2/{\L^\pri}^2)$.  Construction of a suitable $U$ is may
be accomplished following the formulation of Wegner {\cite{wegner}}.
We define the operator $T$ via
\begin{equation}
{dU(\L;\L^\pri)\over d(\L^{-2})} \equiv T(\L)U(\L;\L^\pri)\; .
\end{equation}
The evolution of $H$ under a change in $\L$ is then given by
\begin{equation}
{dH(\L)\over d\L^{-2}} = [T(\L) , H(\L)]\; ,
\label{dhdlambda}
\end{equation}
from which $T$ is seen to be a generator of infinitesimal scale
transformations.  It is straightforward to show that the choice
\begin{equation}
T(\L) = [H_0,H(\L)]
\end{equation}
results in the transformation properly changing the value of the
cut-off parameter in the regulating functions used for all matrix
elements of the Hamiltonian{\cite{bigpaper}}.

Eqn. (\ref{dhdlambda}) can then be analyzed order-by-order in
perturbation theory by expanding the interaction $H_I = \sum_p g^p
H_p$.  At second order, for example, we find
\begin{equation}
\langle n|\d V_2|m\rangle
= \ha\sum_l\left({1\over E_{nl}}+{1\over E_{ml}}\right)
\left(e^{-2E_{nl}E_{ml}/\L^2} - e^{-2E_{nl}E_{ml}/{\L^\pri}^2}\right)
\langle n|V_1(\L)|l\rangle \langle l|V_1(\L)|m\rangle\; ,
\end{equation}
where $\langle n|H_p(\L)|m\rangle \equiv
\exp\left(-E_{nm}^2/\L^2\right) \langle n|V_p(\L)|m\rangle$ defines
the ``reduced'' interaction $V_p$, and $\langle n|\d V_2|m\rangle
\equiv \langle n|V_2(\L^\pri)|m\rangle - \langle n|V_2(\L)|m\rangle$.
This shift in the interaction explicitly accounts for the physics
removed by changing the cut-off from $\L$ to $\L^\pri$.

As discussed above, requiring form-invariance of the Hamiltonian under
the similarity transformation determines the counterterm structure
uniquely.  We have implemented this constraint and determined the
complete effective Hamiltonian through second order.  The operators
that appear in the counterterm Hamiltonian include, as expected, all
possible one-body operators which are rotationally invariant (and
respect, e.g., time-reversal invariance, charge conjugation, etc.).
These are a gluon ``mass'' term, $\bA^2$, the operators $\bPi^2$ and
$(\nabla_i A_j)^2$, and, for fermions, $\psibar\psi$.  Each has a
calculable coefficient, except for the particular combination of
$\bPi^2$ and $(\nabla_i A_j)^2$ which has the structure of a gluon
wavefunction renormalization.  For this combined operator, the {\em
variation} of the coefficient under changes in the cut-off is
calculable, and gives the gauge field anomalous dimension: $\g_A =
-(g^2/16\pi^2)\left(4C_A/3\right)$, where $C_A$ is the Casimir
invariant of the adjoint representation.  See Ref. {\cite{bigpaper}}
for full details.

\section{Work in Progress and Future Directions}

Having obtained the renormalized Hamiltonian, we can now proceed to
construct a model for the ground state and analyze excitations such as
glueballs, mesons and hybrids.  In Ref. {\cite{bigpaper}}, we focused
attention on the gluonic sector and made a variational BCS vacuum {\em
Ansatz}.  This leads to a gap equation which may be solved numerically
to give the spectral function of the quasi-particle constituent
gluons.

Work in progress includes extending the similarity analysis to higher
orders, applying this smooth cut-off treatment to the quark sector and
designing improved models for the vacuum.  At third order, one should
see the running of the coupling, as well as the appearance of many new
operators in the counterterm Hamiltonian.  Going beyond BCS is
expected to be particularly important in a combined treatment of the
quark and gluon sectors.  In addition, other composite operators, such
as currents, must be properly renormalized which is also possible
using the formalism presented in Ref. {\cite{bigpaper}}.

Finally, there are a host of phenomenological applications to be
explored.  Not only meson spectra but also form factors, electro-weak
and hadronic decays, etc. can be predicted and confronted with
available data.  This comprehensive analysis should provide
significant insight into hadron structure as well as a key linkage
between fundamental QCD and the successful phenomenological quark
model.

\section*{Acknowledgments}
Financial support is provided by U. S. Department of Energy.

\end{document}